# Regional Development Classification Model using Decision Tree Approach

Tb. Ai Munandar
Eng. Informatics Dept.
Universitas Serang Raya
Serang – Banten - Indonesia

Edi Winarko
Computer Science and Electronic Dept.
Universitas Gajah Mada
Yogyakarta – Indonesia

## ABSTRACT
Regional development classification is one way to look at differences in levels of development outcomes. Some frequently used methods are the shift share, Gain index, the Iindex Williamson and Klassen typology. The development of science in the field of data mining, offers a new way for regional development data classification. This study discusses how the decision tree is used to classify the level of development based on indicators of regional gross domestic product (GDP). GDP Data Central Java and Banten used in this study. Before the data is entered into the decision tree forming algorithm, both the provincial GDP data are classified using Klassen typology. Three decision tree algorithms, namely J48, NBTRee and REPTree tested in this study using cross-validation evaluation, then selected one of the best performing algorithms. The results show that the J48 has a better accuracy rate which is equal to 85.18% compared to the algorithm NBTRee and REPTree. Testing the model is done to the six districts / municipalities in the province of Banten, and shows that there are two districts / cities are still at the development of the status quadrant relatively underdeveloped regions, namely Kota Tangerang and Kabupaten Tangerang. As for the Central Java Province, Kendal, Magelang, Pemalang, Rembang, Semarang and Wonosobo are an area with a quadrant of development also on the status of the region is relatively underdeveloped. Classification model that has been developed is able to classify the level of development fast and easy to enter data directly into the decision tree is formed. This study can be used as an alternative decision support for policy makers in order to determine the future direction of development.

## 1. INTRODUCTION
Regional development classification activities carried out in order to see the inequality of outcome achievement of development of an area according to specific indicators. Several methods are often used to classify the level of development among others shift share analysis, Gain Index, Index Williamson and typology Klassen [1]. The development of data mining has change many aspect. It is encouraging the scientists to implement it into the process of grouping the level of development according to certain indicators based on clustering techniques. The goal of its, is to see the level of achievement of development disparities between one region to another.

Some of the current research, implement clustering techniques to determine the regional development disparities. Ward cluster technique, for example, used by scientists to determine the inequality of development that occurred in West Germany and East Germany [2], the Czech Republic [3] and Ukraine [4]. Each grouping consecutive region based on economic indicators and Gross Regional Product.

Previous studies using a combination of clustering techniques. Ward and K-means are used to classify the level of development of the region. Where Ward technique is used to form the cluster centroid as the initial centroid of K-means clustering. This inital centroid then used as the $k$ clusters for classification of an area based on its statistical development in K-means [5], [6], [7]. Several other techniques that often used for grouping regions based indicators of achievement level of development are within-group linkage, complete linkage [8] and medoid partitioning algorithm [9]. Although the cluster technique used by some scientists for grouping level of development, on the other hand, there are serious issues of concern to scientists, the outcomes associated with the cluster is still provide great opportunities for subjective interpretation.

The differences grouping level of regional development between cluster results and classification based on government standard rules occurs in [5] and [6]. The study that conducted in [5] and [6] are compared cluster results analysis with the development gap classification according to the European Commission (EC) in the European Union. Their study show that there is a difference between the classification according to EC and cluster results techniques. Its especially indicated by the presence of some regions in a cluster, not in accordance with classification of EC. The exixtence of these differences, of course, provide its own difficulties to correctly interpret the results of the cluster. Difficulties related to the interpretation of determining the disparity of development, also because of the region fall into several clusters at the same time making it difficult to determine exactly, a region into the appropriate cluster or not [10]. Another difficulty due to the cluster technique also does not have a clear label for each group are formed. Whereas, clearly labeled on the results of the cluster is needed to be able to interpret the development of a region-level group [11]. To cope with the difficulties of interpretation, this study propose a decision tree approach to classify the level of development of the region so that the disparity can be seen clearly.

Decision tree is a supervised approach to classify a large number of datasets that make up the structure of the rules are simple, clear and easy to understand [12]. Kalathur (2006) in [13] stated that the decision tree is used to examine the data and form a rules in a tree that will be used for forecasting needs. There are many types of algorithms in the decision tree that can be utilized for a variety of needs [13], some of them and used in this study are J48 which is the development of C4.5, Naive Bayess Tree (NBTree) and Reduced Error Pruning Tree (REPTree).

The main objective of this study is to classify the level of development of a region based on indicators of regional gross domestic product (GDP) using a decision tree. This study will also be tested against three decision tree algorithm before classifying data testing, to choose which algorithm is better so that can be used as a classification tool. Each algorithm will be evaluated by cross-validation technique.

This study is divided into five sections discussion. The first part is an introduction that contains the background issues related to the proposed research topic. The second part contains an overview of related decision tree. The third part is a research methodology that includes information such as the working steps of research, data sources, tools and techniques used. The fourth part is a discussion and discussion of research results. The last part is the conclusion of research conducted.

## 2. KLASSEN TIPOLOGY

Klassen typology is used to provide an overview of the growth pattern of economic development of a region. This technique divide an area into four quadrants based on economic growth and income per capita of the region. The four quadrants of development in question are:

1) Quadrant I, is an advanced and fast-growing area. This category is indicated by the level of economic growth and per capita income of a region, higher than average growth rate compared to the province.

2) Quadrant II, an advanced but pressured area. Indicated by a lower rate of economic growth but per capita income is higher than the provincial average.

3) Quadrant III, is a rapidly growing area. Indicated by a higher rate of economic growth but per capita income is lower than the provincial average.

4) Quadrant IV, is relatively underdeveloped regions, the rate of economic growth and income per capita, lower than the province.

**Tabel 1 Classification of economic growth according Klassen tipology**

| Quadrant I | Quadrant II |
|---|---|
| an advanced and fast-growing area | an advanced but pressured area |
| $r_i > r$ dan $y_i < y$ | $r_i < r$ dan $y_i > y$ |

| Quadrant III | Quadrant IV |
|---|---|
| a rapidly growing area | relatively underdeveloped regions |
| $r_i > r$ dan $y_i > y$ | $r_i < r$ dan $y_i < y$ |

Where $r_i$ is the rate of GDP growth in district $i$, $r$ is the rate of GDP growth in the province, $y_i$ is the average contribution to the development of district $i$ and $y$ is the average contribution to the development of province $i$.

Growth rate is obtained by the equation:

$$r = \frac{Ind\_PDRB_t - Ind\_PDRB_{t-1}}{Ind\_PDRB_{t-1}} x100\% \qquad (1)$$

In this case, *Ind_PDRBt* is an observation of the current GDP indicator, while *Ind_PDRBt-1* is an indicator of GDP a year earlier observations.

The average value of the contribution of the construction is obtained by the equation:

$$y = \frac{Ind\_PDRB_t + Ind\_PDRB_{t-1}}{TI\_PDRB_t + TI\_PDRB_{t-1}} x100\% \qquad (2)$$

Where *TI_PDRBt* is a total value of GDP for all indicators in the observation and *TI_PDRBt-1* is the total value of GDP for all indicators in the previous year. To calculate the growth rate of the district and provincial *(r)* used equation (1), while for calculating the average contribution of development *(y)* use equation (2).

The following illustration is given by Klassen typology grouping calculations to see the development of a district level by comparing the data with the provincial GDP district. Assume that the data used are the 2006 data *(t-1)* and 2007 *(t)*. If the GDP in 2006 for the district is 1,500 and the province is 2,350, then the GDP in 2007 for the district is 1,750 and the province of 2,600. The total value of GDP indicator in 2006 for the district is 14,500 and the province of 25,500 and an indicator total GDP 2007 for the district is 16,500 and the province is 28,900.

Equation (1) provides the value of district growth rate of development *(ri)* amounted to 16.7% and the province *(r)* of 10.6%. The average contribution of development districts (yi) obtained for 10.5% and 9.1% for the province (y) are provided by equation (2). According to Klassen typology rules, it shows that ri > r and yi > y, then it can be obtained that the district enter into areas with rapidly growing development (Quadrant III).

## 3. DECISION TREE

Decision tree is a technique that can be used to classify the data in large size by forming a tree structure which contains the rules that are simple, clear and easy to understand [12]. It can be used for various needs, such as forecasting [13], determination of student exams predictions [14] and the identification of the risk of trauma during delivery [15]. This section, outlines three decision tree algorithm used in this study, namely J48, NBTree and REPTree.

### 3.1 J48 Algorithm

Quinlan (1993) in [16] states that the J48 algorithm is a development of the ID3 algorithm with additional features to overcome the problems that can not be solved in ID3. This algorithm is implemented for many aspects, one of it, is to predict student behavior [13]. Here is the algorithm used in the J48 to build a decision tree.

1) Select the attribute as root

2) Create a branch for each value

3) Perform the division of cases into the branch

4) Repeat the process for each branch until all cases the branches have the same class

These algorithms use the Information Gain and Gain Ratio to determine the main root of the decision tree that will be formed based on an existing attribute.

### 3.2 NBTree Algorithm

This algorithm is used to generate a decision tree from a dataset. Where the determination of leave on rules established based on Naive Bayes technique. This algorithm uses information on the frequency of occurrence of a class. Here is an NBTree algorithm [17]:

1) Define the initial conditions.

2) Group data and then calculate the value of the node that has splited.

3) Prune tree formed to evaluate the optimal tree and cross-validation error

4) Perform the test using test data in a tree and terminal node identification in accordance with its test data.

5) Perform the steps in front of the predictions using Naive Bayes on terminal nodes generated.

## 3.3 REPTree Algorithm

Witten (2005) in [18] suggests that REPTree works on the principle of a calculated gain and entropy information to reduce the occurrence of the error variance of the data. This method can reduce the complexity of the decision tree model by reducing errors during decision tree pruning formed and the result of the variance of the dataset.

## 4. RESEARCH METHODOLOGY

Research steps of this study can be seen in Figure 1. This study begins with the collection of statistical data GDP province of Central Java and Banten. Nine GDP indicators used in this study, including Agriculture, Mining, Industrial, Electrical, Building, Commerce, Transportation, Finance and Services. GDP statistics Central Java province obtained by 135 datasets which then will be used as training data to build classification model and vice versa. This is applies also for 54 datasets of Banten GDP.

To obtain training data (Central Java; Banten) and data testing (Banten;Central Java), of the nine indicators of GDP growth and contribution calculated value both district and provincial development in order to obtain the four attributes predictor produces a single class classification using Klassen tipology.

The next step is testing the training data using the three algorithms, namely J48, NBTree and REPTree. All three algorithms were evaluated using cross-validation technique. They generate decision trees and then compared to select proper methods with the highest accuracy. Measurement comparison method performed by looking at the value of their accuracy, kappa, MAE and MSE. The comparison then choose the method with the best results. The method chosen then tested using the data testing (Banten), and then compared with the data back to the data grouping Klassen Banten province.

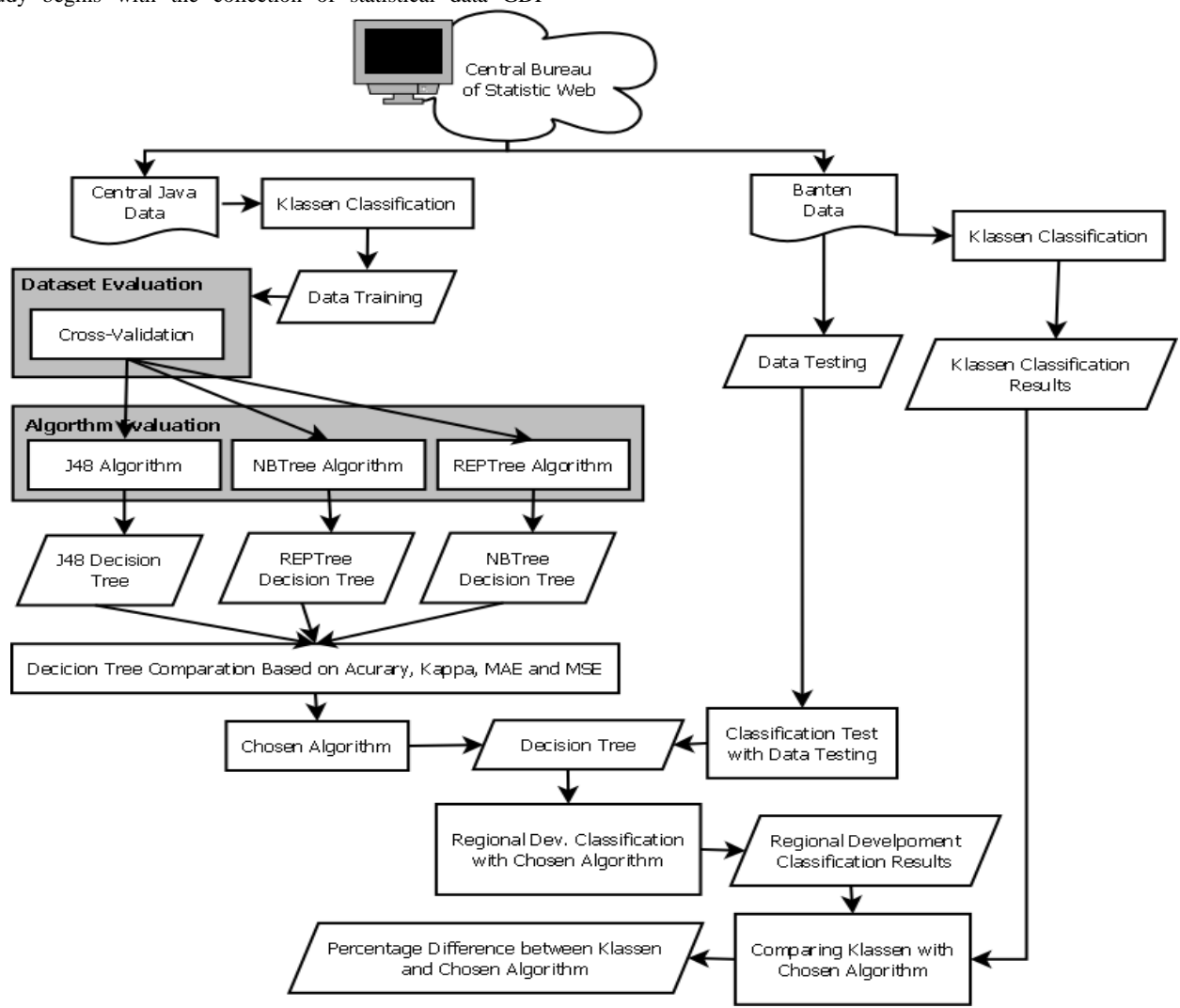

**Fig 1: Research steps**

## 5. RESULT AND DISCUSION
### 5.1 Comparative Analysis of Algorithms

Table 2 gives the results of the grouping level of development of the region according to the Klassen. It shows that the level of development in Banten province tend to be in Quadrant II, III and IV, while Central Java province, dominated by Quadrant I, II and IV.

**Tabel 2 Results of grouping level of development and Banten Central Java typology based**

| Klassen Quadrant | Province | |
|---|---|---|
| | Banten | Jawa Tengah |
| I | 6 | 23 |
| II | 19 | 48 |
| III | 15 | 17 |
| IV | 14 | 47 |

Training data was tested using Knowledge Flow tool via Weka application. It was tested by three algorithms (J48, NBTree and REPTree) and validated via corss-validation with a number of fold = 10 to see the level of accuracy that is owned by the three algorithms.

Table 3 shows the three algorithms results test. J48 algorithm with cross-validation has the highest Kappa value which is equal to 0.6989. This means that the variable conformity with J48 testing better than other algorithms, although it only has a slight difference in the Kappa value.

**Tabel 3 Comparison of test results J48, NB-Tree and REP-Tree with Cross-validation**

| Algorithm | J48 | NBTree | REPTree |
|---|---|---|---|
| Classification accuracy (%) | 85.18 | 82.22 | 84.44 |
| Kappa | 0.699 | 0.615 | 0.661 |
| Mean absolute Error | 0.072 | 0.134 | 0.124 |
| Root mean squared error | 0.248 | 0.271 | 0.265 |

Measurement accuracy is more, performed with a great look deviation (error) prediction results indicated by means absolute error (MAE) and root mean squared error (MSE). The smaller the deviation value, the better the performance of the algorithm. The results showed that, the value of MAE and MSE of J48 algorithm, smaller than NBTree algorithms and REPTree. Thus it can be said that the J48 <NBTree <REPTree.

Classification accuracy for each algorithm, also put J48 as algorithms on a higher degree of accuracy than the others. J48 algorithm has a classification accuracy rate of 85.18%, NB-Tree by 82.22% while the REP-Tree of 84.44%.

The decision tree results also show that REPTree eliminate class K3 and NBTree algorithm is only able to recognize the class K1 and K4. While J48 able to recognize all class (K1, K2, K3, and K4). Thus, the model developed, J48 algorithm chosen as the algorithm for classifying the level of development of the region in accordance with its GDP indicator.

### 5.2 Model Testing

This section discusses the testing of models created by the algorithm selected in section 5.1. A total of 54 data is testing (Banten) and 135 data testing (Central Java) were tested using the rules of the decision tree of J48. Fifty-four of data testing represent the six districts in the province of Banten. These data testing also become data training for testing the model later.

**Tabel 4 the level of accuracy J48 after pruning and unpruned**

| | P1 | P2 | P3 | P4 |
|---|---|---|---|---|
| Pruning | 25.93% | 3.70% | 50.00% | 67.41% |
| Un-pruned | 33.33% | 65.19% | 51.85% | 82.96% |

At this study, four test groups (P1, P2, P3 and P4) was conducted to test the model of development level classifier using J48. Where P1 shows the tests performed by the Central Java data as training data, then Banten as a data testing. P2 shows that Banten as data training, and Central Java as a data testing. P3 shows the tests performed where Banten and Central Java incorporated as training data, then take the data Banten as a data testing. While P4 shows the same test performed at P3, only, testing the data used is Central Java. This study was conducted either by pruning techniques and non-pruning on the decision tree is formed. Table 4 above shows a comparison test using J48 classification model with different conditions (P1, P2, P3 and P4).

Test results P1 non-pruning shows that there are 18 data which have the same class with Klassen classification. It means that the results of the predictive classification rules J48 has an accuracy rate of 33.33% when compared to the grouping using Klassen typology. While at P1 with prunning testing, classification accuracy rate rose to 25.93%. At P1 testing both pruning and non-prunning indicates that quadrant development using J48 classification tend to be in Quadrant 2 (see Figure 2).

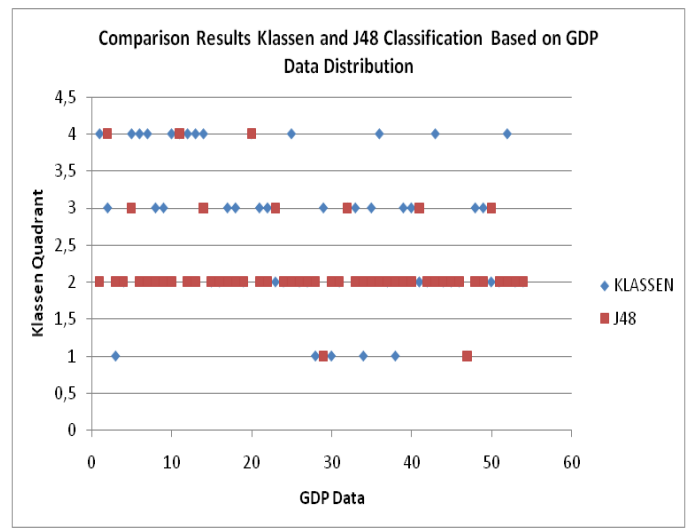

**Fig 2: Comparison of typology Klassen and J48 according to the data distribution**

Testing P2 indicates the level of classification accuracy of 65.19% prior to the pruning of the decision tree is formed. Meanwhile, after the pruning of the decision tree, the results of the classification accuracy rate has decreased significantly to 3.70%. On testing P2 is also seen that the level of development of the region based on the pruning process is dominated by quadrant III (K3), while for non-prunning tend to be in quadrant III (K3) and IV (K4).

Table 5 shows the quadrant level of development of an area based on all tests performed, either by pruning and non-pruning of the decision tree.

**Tabel 5 Quadrant development to GDP indicator based classification J48 with testing P1, P2, P3 and P4**

| Quadrant | Non Tree Pruning | | | | Tree Pruning | | | |
|---|---|---|---|---|---|---|---|---|
| | P1 | P2 | P3 | P4 | P1 | P2 | P3 | P4 |
| I | 10 | 0 | 7 | 15 | 20 | 0 | 5 | 6 |
| II | 11 | 8 | 19 | 39 | 15 | 8 | 17 | 48 |
| III | 11 | 31 | 5 | 21 | 10 | 127 | 14 | 29 |
| IV | 22 | 96 | 23 | 60 | 9 | 0 | 18 | 52 |

Tables 6 and 7 show the results of the classification level of development by the district / city using a combination of data trainining tested to the province, both Banten and Central Java.

**Tabel 6 Quadrant development to GDP indicator based classification J48 for Banten**

| DISTRICT | Class | | |
|---|---|---|---|
| | Klassen | J48-Pruning | J48-No-Pruning |
| KOTA CILEGON | K2 | K3 | K4 |
| KOTA TANGERANG | K2 | K4 | K4 |
| KOTA SERANG | K2 | K2 | K2 |
| KAB. TANGERANG | K3 | K4 | K2 |
| KAB. LEBAK | K2 | K2 | K4 |
| KAB. PANDEGLANG | K3 | K2 | K2 |

**Tabel 7 Quadrant development to GDP indicator based classification J48 for Central Java Province**

| DISTRICT | Class | | |
|---|---|---|---|
| | Klassen | J48-Pruning | J48-No-Pruning |
| BANYUMAS | K2 | K2 | K2 |
| BLORA | K2 | K2 | K4 |
| CILACAP | K2 | K2 | K2 |
| DEMAK | K3 | K3 | K3 |
| KUDUS | K3 | K3 | K4 |
| KENDAL | K2 | K4 | K4 |
| KOTA SALATIGA | K2 | K2 | K2 |
| MAGELANG KAB | K2 | K4 | K4 |
| PEKALONGAN | K2 | K2 | K4 |
| PEMALANG | K2 | K4 | K2 |
| PURBALINGGA | K2 | K3 | K4 |
| REMBANG | K2 | K4 | K2 |
| SEMARANG KAB | K2 | K4 | K4 |
| TEMANGGUNG KAB | K2 | K2 | K4 |
| WONOSOBO KAB | K2 | K4 | K4 |

Where K1 is an advanced and fast growing area, K2 was developed areas but depressed, K3 is a rapidly growing area while K4 is relatively underdeveloped area.

Table 6 above shows that the city of Tangerang and Tangerang, are still relatively underdeveloped regions quadrant when the analysis using a model with a J48 decision tree pruning. As for the Central Java province, there are five districts / cities are still in the quadrant with establishment of the status of the relatively underdeveloped regions, namely Kendal, Magelang, Pemalang, Rembang, Semarang and Wonosobo.

Overall, the results showed that the classification level of development using J48 classification technique provides results that are not the same as the Klassen typology. This is because Klassen typology classifies the GDP data of an area based on mathematical calculations, while the J48, extract tendency possessed of a data pattern to form a classification rule. Another advantage of this classification technique is, any data can be classified into classes according to the rules that have been clearly established.

The analysis conducted in this study may generally be used as a tool to determine the future direction of development of the region. Development carried by looking at the level of performance achieved by an area. So that the development process more effective.

## 6. CONCLUSION
Based on the first study conducted, which is to compare the three algorithms, J48 algorithm has an accuracy rate of classification rules formation better than REPTree and NBTree algorithm, which amounted to 85.18%. Suitability of predictor variables on the results of the decision tree formed from J48 has a better level of fitness and have a deviation value prediction results are smaller than the REPTree and NBTree. Decision tree formed from J48 algorithm, can be used as a rule of classification levels by utilizing the regional development and the growth rate variable contribution of both district and provincial development taken GDP indicator.

The results of classification model shows that there are two districts in Banten that still in a relatively underdeveloped area status, namely Kota Tangerang and Kabupaten Tangerang. While, Kendal, Magelang, Pemalang, Rembang, Semarang and Wonosobo are an area in Central Java with a quadrant of development also on the status of the region is relatively underdeveloped.

This study can be used as an alternative decision support for policy makers in order to determine the future direction of development. Future studies will be directed to the development of a group decision support application to assess the level of development of the region.